\documentclass[prc,aps,groupedaddress,preprintnumbers]{revtex4}
\usepackage{graphicx} 
\begin{document}

\title{Relation between the $2\nu\beta\beta$ and $0\nu\beta\beta$ nuclear matrix 
elements}

\author{Petr Vogel}
\email{pxv@caltech.edu}
\affiliation{\it Kellogg Radiation Laboratory, Caltech, Pasadena, CA 91125, USA}
\author{Fedor \v{S}imkovic}
\email[]{fedor.simkovic@fmph.uniba.sk}
\affiliation {\it Department of Nuclear Physics and Biophysics, 
Comenius University,
Mlynsk\'{a} dolina F1, SK-84248 Bratislava, Slovakia}

\begin{abstract}
A formal relation between the GT part of the nuclear matrix elements $M^{0\nu}_{GT}$
of $0\nu\beta\beta$ decay and the closure matrix elements $M^{2\nu}_{cl}$ of 
$2\nu\beta\beta$ decay is established. This relation is based on the integral representation
of these quantities in terms of their dependence on the distance $r$ between the 
two nucleons undergoing transformation. We also discuss the difficulties in determining
the correct values of the closure $2\nu\beta\beta$ decay matrix elements.
\end{abstract}

\maketitle


\section{Introduction}

The $2\nu\beta\beta$ decay mode of  most candidate nuclei has been observed
 and its half-life determined. Thus, the corresponding nuclear matrix elements
$M^{2\nu}$ are known. These matrix elements, of dimension energy$^{-1}$, vary
abruptly between nuclei with different $Z$ and $A$; they exhibit pronounced shell effects.
In contrast, the fundamentally more important $0\nu\beta\beta$ decay mode
have not been reliably observed so far. Therefore, the corresponding dimensionless nuclear 
matrix elements $M^{0\nu}$ must be evaluated theoretically. These calculated quantities,
whether based on the QRPA \cite{us1,us2,us3,SC}, nuclear shell model
\cite{Men08,Cau08,Cau08a}, or the Interacting Boson Model \cite{Bar09}, do not show
such a variability; instead they vary relatively smoothly between nuclei with different $Z$,
and $A$. Evaluation of the $M^{0\nu}$ using the Generator Coordinate Method \cite{Gabriel} or
the Projected Hartree-Fock-Bogolyubov Method \cite{PHFB} vary smoothly with
$Z$ and $A$ as well.

However, the calculated values of the matrix elements $M^{0\nu}$ using different approximations
do not agree with each other perfectly, differences of a factor of about two exist. 
Moreover, given the fundamental importance of these quantities for the planning and 
interpreting the $0\nu\beta\beta$
decay experiments, it would be good to have independent observables that could be linked to
their magnitude. 
In that context we wish to address in this work several questions:
\begin{itemize}
\item Can one understand 
intuitively the different behavior of  $M^{0\nu}$ and  $M^{2\nu}$ when
of $Z$ and $A$ are varied ?
\item Is there a formal relation between  $M^{0\nu}$ and  $M^{2\nu}$ ?
\item If such relation exists can it be used to test the calculated values of the $0\nu\beta\beta$
matrix elements ?
\end{itemize}  

In discussing these problems we basically follow the work published
by us and our collaborators  earlier in Ref. \cite{us4}.

\section{Formalism}

Assuming that the  $0\nu\beta\beta$ decay is caused by exchange of the light
 Majorana neutrinos, the nuclear matrix element  consists of 
Gamow-Teller, Fermi and Tensor 
parts,         
\begin{equation}
 M^{0\nu} = M^{0\nu}_{GT} - \frac{M^{0\nu}_F}{g_A^2} +  M^{0\nu}_T \equiv  
M^{0\nu}_{GT} ( 1 + \chi_F + \chi_T ) ~,
\end{equation}
where $ \chi_F$ and $ \chi_T$ are matrix element ratios that are smaller than unity and, 
presumably, less dependent on the details of the applied nuclear model.

In the following we concentrate on the GT part, $ M^{0\nu}_{GT} $, which can be somewhat 
symbolically written as
\begin{equation}
 M^{0\nu}_{GT} = \langle f | \Sigma_{lk} \vec{\sigma}_l \cdot  \vec{\sigma}_k \tau_l^+ \tau_k^+
H(r_{lk},\bar{E}) | i \rangle ~,
\end{equation}
where $ H(r_{lk},\bar{E})$ is the neutrino potential and $r_{lk}$ is the relative
distance between the two neutrons that are transformed in the decay into the two protons.

The dependence of $M^{0\nu}$ on the distance $r_{lk}$ is described by the function 
$C^{0\nu}(r)$ (first introduced in \cite{jon} see also \cite{us3,Men08})
\begin{equation}
C^{0\nu}_{GT}(r) =  
\langle f | \Sigma_{lk} \vec{\sigma}_l \cdot  \vec{\sigma}_k \tau_l^+ \tau_k^+
\delta(r - r_{lk}) H(r_{lk},\bar{E}) | i \rangle ~,~~~
 M^{0\nu}_{GT} = \int_0^{\infty} C^{0\nu}_{GT} (r) dr ~,
\label{eq:C(r)}
\end{equation}
where $\delta(x)$ is the Dirac delta function.

In analogy to Eq. (\ref{eq:C(r)}) we can define for the case of the $2\nu\beta\beta$ decay
a new function
\begin{equation}
C^{2\nu}_{cl}(r) =  \langle f | \Sigma_{lk} \vec{\sigma}_l \cdot  \vec{\sigma}_k
\delta(r - r_{lk}) \tau_l^+ \tau_k^+ | i \rangle ~, ~~~~
M^{2\nu}_{cl} = \int_0^{\infty} C^{2\nu}_{cl}(r) dr ~.
\label{eq:2nuclp}
\end{equation}
This function, therefore, is related to the 
dimensionless {\it closure} matrix element for the
$2\nu\beta\beta$ decay, not the true, dimension energy$^{-1}$ matrix element
$M^{2\nu}$ that contains the corresponding energy denominators.
While the matrix elements $M^{2\nu}$ and $M^{2\nu}_{cl}$ get contributions only from the $1^+$
intermediate states, the function $C^{2\nu}_{cl}(r)$ gets contributions
from all intermediate multipoles.
This is the consequence of the $\delta$ function in the definition of $C^{2\nu}_{cl}(r)$.
Naturally, when integrated over $r$ only the contributions
from the $1^+$ are nonvanishing.

It is now clear that, by construction,
\begin{equation}
C^{0\nu}_{GT}(r) =  H(r,\bar{E}) \times C^{2\nu}_{cl}(r) ~,
\label{eq:basic}
\end{equation}
which is valid for any shape of the neutrino potential
$ H(r,\bar{E})$. Thus, if $C^{2\nu}_{cl}(r)$ is known,
$C^{0\nu}_{GT}(r)$ and therefore also $M^{0\nu}_{GT}$ can be easily determined
since the neutrino potential $H(r,\bar{E})$ is known and only
weakly dependent on the average excitation energy $\bar{E}$.
The equation (\ref{eq:basic}) represents the basic
relation between the $0\nu$ and $2\nu$ $\beta\beta$-decay modes.

\begin{figure}
  \includegraphics[height=.37\textheight]{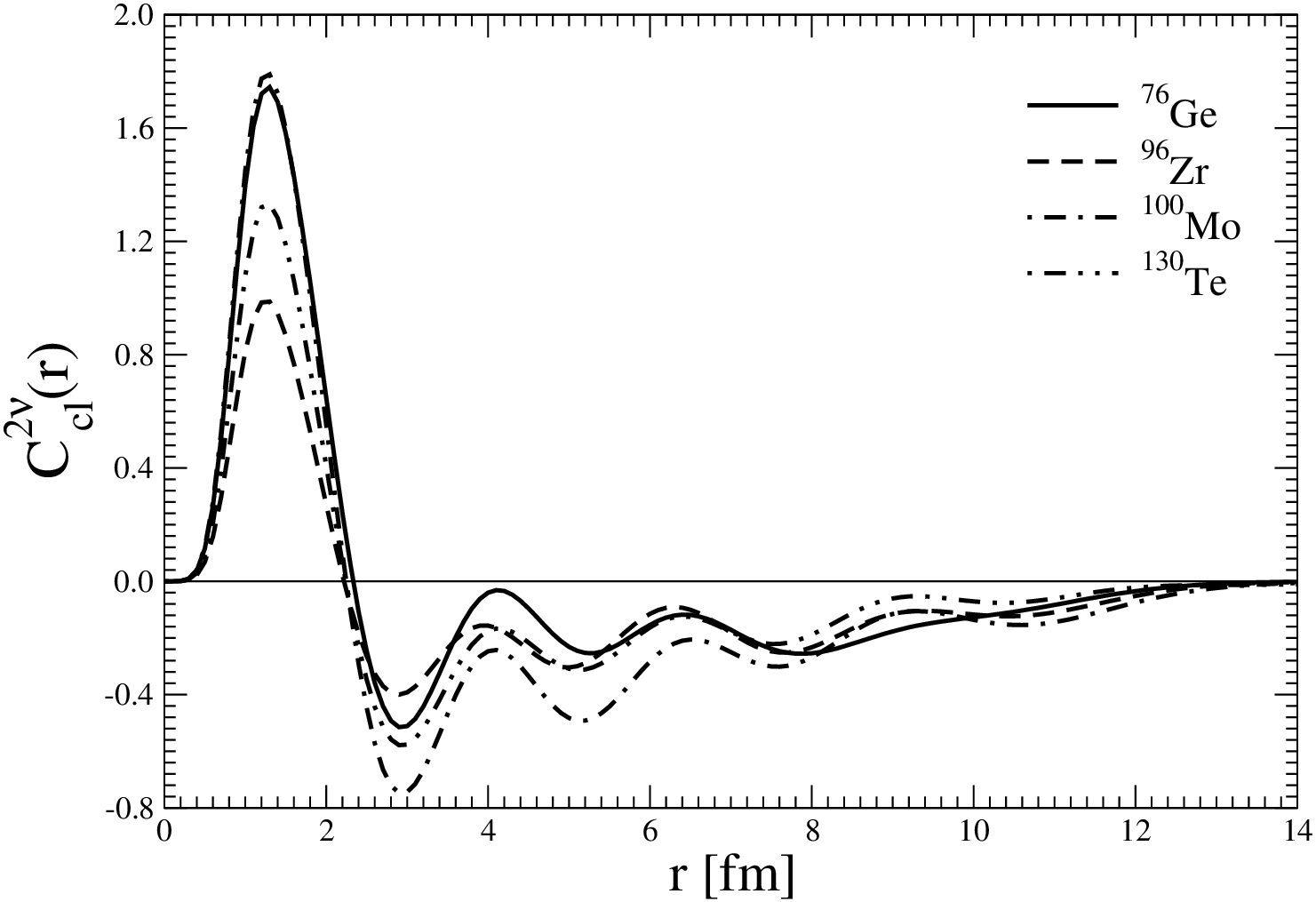}
  \caption{The function $C^{2\nu}_{cl}(r)$ for different selected candidate nuclei.} 
\label{fig:C(r)}
\end{figure}
Examples of the function  $C^{2\nu}_{cl}(r)$ are shown in Fig. \ref{fig:C(r)} .
Note that while the function $C^{2\nu}_{cl}(r)$ has a substantial negative tail 
past $r \sim 2-3$ fm,
these distances contribute very little to $C^{0\nu}_{GT}(r)$ and, therefore also
to $M^{0\nu}$. This is a consequence of the
shape of the neutrino potential $H(r,\bar{E})$ that decreases fast
with increasing values of the distance $r$.

\section{Results and discussion}

Remembering that in a nucleus the average distance between nucleons is $\sim$1.2 fm
we can somewhat schematically separate the range of the variable $r$ in the functions
$C^{0\nu}_{GT}(r)$ and $C^{2\nu}_{cl}(r)$  into the region $r \le$ 2-3 fm governed
by the nucleon-nucleon correlations, while the region $r \ge$ 2-3 fm is governed by
nuclear many-body physics. From the form of $C^{0\nu}_{GT}(r)$ it follows that
the matrix elements $M^{0\nu}_{GT}$ are almost independent of the ``nuclear"
region of $r$ and  hence one does not expect rapid variations of their value when $A$
or $Z$ of the nucleus is changed. On the other hand, the $2\nu$ closure matrix elements
$M^{2\nu}_{cl}$ depend sensitively on that region of $r$ 
since there is a substantial cancellation between the positive part at $r \le 2-3$ fm 
and the negative tail at $r \ge 2-3$ fm. Hence one expects sizable shell effects,
i.e. a significant variation of $M^{2\nu}$ and $M^{2\nu}_{cl}$ with $A$ and $Z$,
in agreement with observations. We have thus answered the first two questions posed
in the Introduction.

But answering the third question is not so simple.
While the nuclear matrix elements $M^{2\nu}$ are simply related to the $2\nu$ half-life
$T^{2\nu}_{1/2}$, and are thus  known for the nuclei in which $T^{2\nu}_{1/2}$ has
been measured, the closure matrix elements  $M^{2\nu}_{cl}$ need be determined separately.
One can rely on a nuclear model (e.g. QRPA or nuclear shell model), adjust parameters
in such a way that the experimental value of $M^{2\nu}$ is correctly reproduced,
and use the model to evaluate $M^{2\nu}_{cl}$. Alternatively, one could use the measured 
$\beta^-$ and $\beta^+$ strength functions and assume
coherence (i.e. same signs) among states with noticeable strengths in both channels.
In this way an upper limit of  $M^{2\nu}_{cl}$ can be obtained. Obviously, neither of these
procedures is exact. In the following we leave temporarily aside the important
question to which extent the function $C^{2\nu}_{cl}(r)$ is 
strongly constrained by its integral
value $M^{2\nu}_{cl}$. 

We have shown in Ref. \cite{us4} that, within QRPA, the quantities $M^{2\nu}_{cl}$ are
negative in many cases, contrary to simple expectations (because one obviously expects
that the average excitation energy is positive). On the other hand, the second method,
mentioned above,
leads to the positive  $M^{2\nu}_{cl}$ by definition. Thus, the two methods disagree
with each other.

\begin{figure}
  \includegraphics[height=.37\textheight]{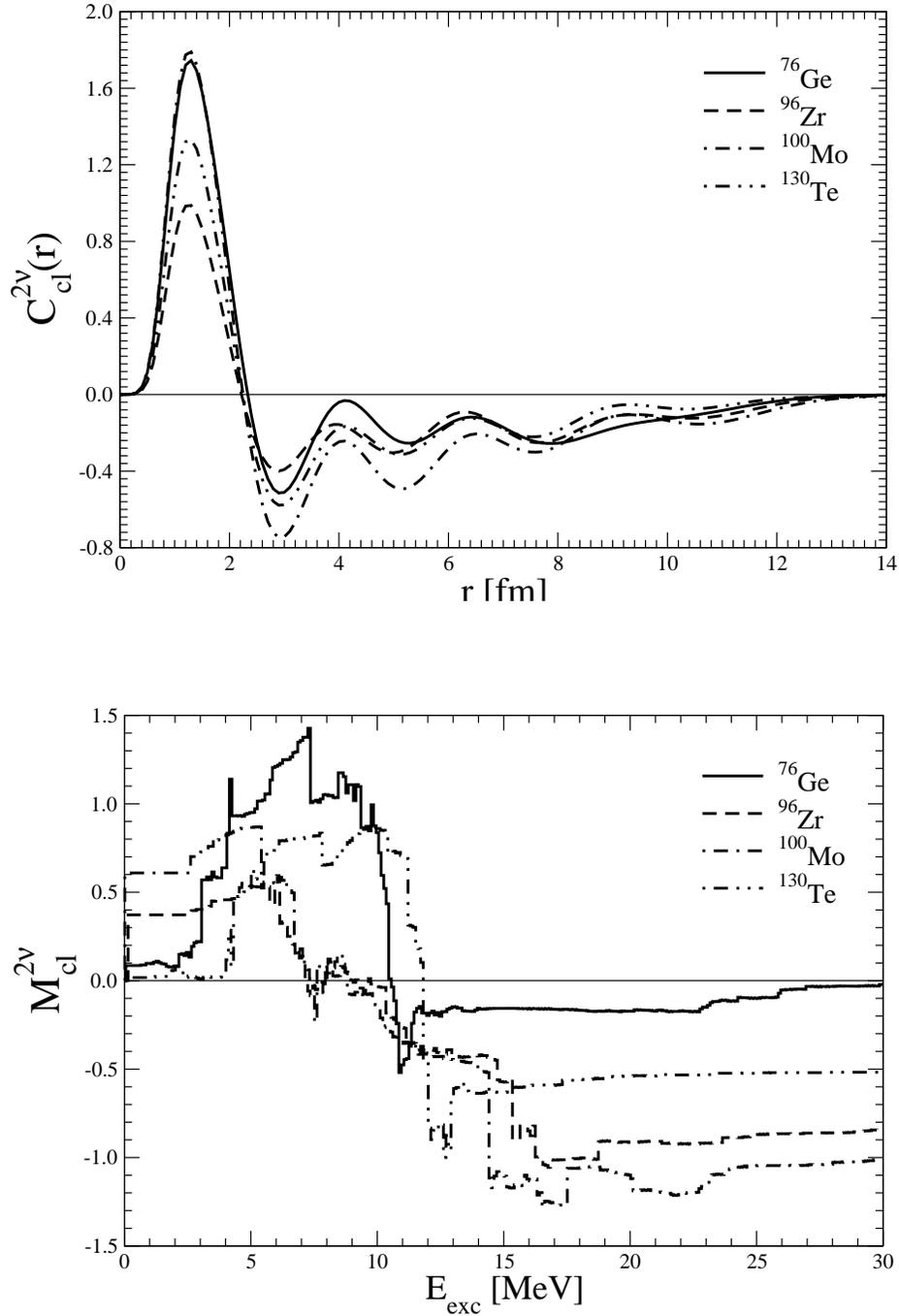}
  \caption{Running sums of $M^{2\nu}_{cl}$ for different selected nuclei as a function
of the excitation energy of the $1^+$ states in the intermediate odd-odd nucleus.}
\label{fig:stairs}
\end{figure}

Since there is a substantial experimental activity devoted to the determination of the
$\beta^{\pm}$ strengths \cite{betas}, it is worthwhile to examine in more detail the somewhat unexpected
finding that in many cases $M^{2\nu}$ and $M^{2\nu}_{cl}$ have opposite signs. Obviously,
this has to do with the different weights of the individual $1^+$ intermediate states
in  $M^{2\nu}$ and $M^{2\nu}_{cl}$. Because of this the higher energy
excited $1^+$ states contribute 
substantially more to $M^{2\nu}_{cl}$  than to  $M^{2\nu}$.  
In Fig. \ref{fig:stairs} we plot the  running sums of  $M^{2\nu}_{cl}$ as the function
of excitation energy in the intermediate nucleus.
One can see that there are negative contributions to the $M^{2\nu}_{cl}$ values 
that  arise from excitation energies
$E_{ex} \ge 10$ MeV where it is difficult to explore experimentally
the corresponding $\beta^+$ strength. Note, moreover,
that this is the region of the giant GT resonance, with substantial $\beta^-$ strength.
Hence even a relatively small $\beta^+$ strength can have a sizable effect.

The negative contribution to  $M^{2\nu}$ and $M^{2\nu}_{cl}$ from relatively high
lying $1^+$ states seems to be a generic effect present in essentially all nuclei.
(Note that analogous effect is present in the shell model evaluation of $M^{2\nu}$
for $^{48}$Ca, see Ref. \cite{Horoi}, Fig. 1.). Thus, it appears that in many cases
when the matrix elements $M^{2\nu}$ (with energy denominators, see \cite{us4})
are evaluated as a function of the intermediate
state excitation energy, the final correct value is reached twice; once at relatively
low excitation energy and the second time asymptotically. 

These negative contributions to  $M^{2\nu}$ and $M^{2\nu}_{cl}$, coming from the
vicinity of the giant GT state, clearly exist only in calculations that are able
to describe the giant GT resonance. In the restricted shell model calculations
(without the full set of the spin-orbit partners) such high lying $1^+$ will be absent.
Thus, we need to make sure that a noticeable $\beta^+$ strength connecting the
final nucleus with the $1^+$ states in the vicinity of the GT resonance really exists
and is not an artifact of our QRPA evaluation. Unless and until this dilemma is resolved 
it is premature to
proceed further in our original program of finding connection
between the  nuclear matrix elements of the $0\nu$ and $2\nu$ $\beta\beta$-decay
modes.

Nevertheless, we have established a formal relation between the nuclear matrix elements
$M^{0\nu}_{GT}$ of the neutrinoless $\beta\beta$ decay and the $M^{2\nu}_{cl}$, the closure
matrix element for the $2\nu\beta\beta$ decay. We also pose a challenge to both experimentalists
studying the charge exchange reactions of the $(n,p)$ type, and to theorists using 
methods  alternative to QRPA to establish whether a noticeable $\beta^+$ strength at
energies near the giant GT resonance exists or not.


\section*{Acknowledgments}
The authors would like to thank the organizers of MEDEX'11 for an opportunity
to present their result.
The work of P.V.\ was
partially supported by the US Department of Energy under Contract No.\
DE-FG02-88ER40397.  F.\v S acknowledges
the support by the VEGA Grant agency  under the contract No.~1/0249/03.


\end{document}